\documentstyle[prc,preprint,aps]{revtex}
\begin{document}
\draft
\title{Demonstration of the auxiliary-field Monte Carlo approach\\
for $\bbox{sd}$-shell nuclei}
\author{W.~E. Ormand, D.~J. Dean, C.~W.~Johnson\cite{CWJAuth},
G.~H.~Lang,
and S.~E.~Koonin}
\address{W.~K.~Kellogg Radiation Laboratory, 106-38\\
California Institute of Technology,\\
Pasadena, CA 91125}
\maketitle
\begin{abstract}
We apply the auxiliary-field Monte Carlo approach to the nuclear
shell model in the $1s$-$0d$ configuration space. The Hamiltonian was
chosen to have isovector pairing and isoscalar multipole-multipole
interactions, and the calculations were performed within the
fixed-particle, canonical ensemble. The results demonstrate the
feasibility of the method for $N\neq Z$ even-even and odd-odd $N=Z$
nuclei. In particular, static observables for even-even Ne isotopes
and ${}^{22}$Na compare well with results obtained from exact
diagonalization of the Hamiltonian. Response functions are presented
for ${}^{22}$Ne and compared with exact results, and the viability of
cranked calculations for $N\neq Z$ even-even nuclei is addressed. We
present methods for computing observables in the canonical ensemble
using Fourier extraction, and for determining the nuclear shape.
\end{abstract}

\section{Introduction}

The shell model is one of the most successful descriptions of
many-fermion systems \cite{ref1}. In this picture, valence particles
are spatially confined by a one-body potential and influence each
other via a residual two-body interaction. Mathematically, the shell
model can be reduced to a matrix-diagonalization problem by computing
the matrix elements of the Hamiltonian between a set of basis states
that span the configuration space of interest. There are several
computer programs that implement this approach (eg.
Ref.~\cite{oxbash}), and impressive agreement between theoretical
calculations and experimental data has been achieved for nuclei with
$A\leq40$ \cite{ref2,ref3}.

The shell-model approach is limited primarily by the combinatorial
growth in the number of basis states with both the number of valence
particles ($N_v$) and the size of the single-particle basis ($N_s$).
Indeed, for nuclei with $A \sim 60$, an unrestricted shell-model
calculation utilizing the $0f_{7/2}$-$0f_{5/2}$-$1p_{3/2}$-$1p_{1/2}$
orbits would involve approximately $2 \times 10^9$ basis states with
definite $z$-projection of angular momentum \cite{ref4}. Angular
momentum and isospin projection would reduce the size of the basis to
about $10^7$. Clearly, a problem of this magnitude lies beyond the
capability of today's computers.

The traditional approach to circumvent the computational limitations
inherent in the shell model is to impose what are often severe and
{\it ad hoc} truncations on the number of basis states.
Unfortunately, because of the strong character of the residual
interaction, calculations of this nature can be unreliable, and
significant renormalizations of the residual interaction and
transition operators are required.

In two previous papers \cite{ref5,ref6}, we presented a Monte Carlo
method that can give an exact treatment of a nuclear shell-model
Hamiltonian, $H$, even in situations where the matrix-diagonalization
technique is impractical. It is based on using the imaginary-time
propagator $\hat U=\exp(-\beta \hat H)$ to either perform a
thermodynamical trace (canonical or grand canonical) at a temperature
$T=\beta^{-1}$ or, for large $\beta$, to filter a many-body trial
state to the exact ground state. By applying the Hubbard-Stratonovich
transformation \cite{ref7}, the two-body terms in $U$ are linearized
with the introduction of an auxiliary field, and the problem is
reduced to a multi-dimensional integral whose dimensions scale more
gently with either $N_v$ or $N_s$ \cite{ref5,ref6}.

In Ref.~\cite{ref5}, the Monte Carlo method was demonstrated with
selected results for $N=Z$ even-even nuclei in the $1s$-$0d$- and
$0f$-$1p$-shells, while much of the formalism for Monte Carlo
approaches to the nuclear shell model is given in detail in
Ref.~\cite{ref6}. The purpose of this work is to further demonstrate
the feasibility of the technique by presenting systematic results for
$N\neq Z$ even-even nuclei, as well as $N=Z$ odd-odd nuclei in the
$1s$-$0d$ shell-model space. In addition, we present a Fourier method
for evaluating the canonical trace that avoids some of the numerical
problems of the activity expansion method presented in
Ref.~\cite{ref6}. We also present a method for deducing the nuclear
shape from our calculations.

We begin with a very brief description of the formalism for Monte
Carlo techniques in Section~2; Subsection~2.1 describes the new
Fourier method for computing the canonical trace. In Section~3,
results obtained for even Ne isotopes (including cranking) and
${}^{22}$Na are given, and in Section~4, a method for deducing the
nuclear shape is described and applied to these calculations.

\section{Formalism}

In this section, we briefly describe the Monte Carlo approach to the
nuclear shell model utilizing auxiliary fields, referring the reader
to Ref.~\cite{ref6} for more details.

In this work, we will consider only the thermal formalism, and so
consider the partition function
\begin{eqnarray}
Z=\hat{\rm T}{\rm r}\exp(-\beta \hat H),
\label{part}
\end{eqnarray}
where $\hat H$ is a generalized Hamiltonian for the system (which may
include Lagrange multipliers constraining particle number or the
$z$-projection of the angular momentum), $\beta$ may be interpreted
as an inverse temperature ($T=1/\beta$), and $\hat{\rm T}{\rm r}$
represents either a grand-canonical trace over all many-body states
in the space (denoted by $\hat{\rm T}{\rm r}_G$) or a canonical trace
with fixed particle number $N_v$ (denoted by $\hat{\rm T}{\rm
r}_{N_v}$). Given the partition function in Eq.~(\ref{part}), the
thermal observable of an operator $\hat O$ is
\begin{eqnarray}
\langle\hat{O}\rangle=
\frac{1}{Z}\hat{\rm T}{\rm r}\left[\hat{O}\exp(-\beta \hat H)\right].
\label{obs}
\end{eqnarray}

We restrict ourselves to Hamiltonians that contain at most two-body
terms such that
\begin{eqnarray}
\hat H=\sum_\alpha \epsilon_\alpha a^\dagger_\alpha a_\alpha +
\frac{1}{2}\sum_{\alpha\beta\gamma\delta}
V_{\alpha\beta\gamma\delta} a^\dagger_\alpha a^\dagger_\beta
a_\delta a_\gamma\;,
\label{Ham}
\end{eqnarray}
where $a^\dagger_\alpha$ and $a_\alpha$ are the anti-commuting
creation and annihilation operators associated with the
single-particle state $\alpha$ defined by the complete set of quantum
numbers $nljmt_z$ ($n$, $l$, $j$, $m$, and $t_z$ denote the
principal, orbital angular momentum, total single-particle angular
momentum, $z$-projection of $j$, and the third component of the
isospin quantum numbers, respectively), and $\epsilon_\alpha$ and
$V_{\alpha\beta\gamma\delta}$ are the single-particle energies and
two-body matrix elements of the residual interaction. Given any
two-body Hamiltonian of the form Eq.~(\ref{Ham}), it is possible to
find a convenient set of one-body operators $\hat {\cal O}_\alpha$ so
that $\hat{H}$ may be written as
\begin{eqnarray}
\hat H=\sum_\alpha \epsilon_\alpha \hat {\cal O}_\alpha+
\frac{1}{2}\sum_\alpha V_\alpha \hat {\cal O}_\alpha^2\;.
\end{eqnarray}

To simplify the imaginary-time propagator $\hat U$, the
Hubbard-Stratonovich transformation makes use of the identity
\begin{eqnarray}
e^{\frac{1}{2}\Lambda {\hat {\cal O}}^2} =
\sqrt{\frac{|\Lambda |}{\pi}} \int d\sigma
e^{-\frac{1}{2}|\Lambda |\sigma^2 + s\Lambda {\hat {\cal O}}}\;,
\label{HS}
\end{eqnarray}
where $s=\pm 1$ if $\Lambda \geq 0$ or $\pm i$ if $\Lambda < 0$. In
general, the $\hat {\cal O}_\alpha$ do not commute, and when
Eq.~(\ref{HS}) is applied to $\hat{U}$, the resulting integral is
accurate only to order $ \beta V_\alpha {\hat {\cal O}}_\alpha^2$.
The accuracy is improved by dividing $\hat U$ into $N_t$ ``time
slices'', so that $\hat U = \left ( \exp[-\Delta \beta \hat H] \right
)^{N_t}$, and applying Eq.~(\ref{HS}) to each ``slice''. For what
follows, we introduce $\tau_n$ as an imaginary time in the range
$(0,\beta)$ defined as $\tau_n= n\Delta\beta$, with $\Delta\beta =
\beta/N_t$.

The thermal observable in Eq.~(\ref{obs}) can now be written as
\begin{eqnarray}
\langle\hat{O}\rangle=
\frac{\int{\cal D}[\sigma]{\cal G}(\sigma)\langle \hat{O}
\rangle_\sigma
\zeta(\sigma)} {\int{\cal D}[\sigma]{\cal G}(\sigma)\zeta(\sigma)},
\label{mceq}
\end{eqnarray}
where ${\cal D}[\sigma] = \prod_{n,\alpha} d\sigma_\alpha (\tau_n)$
is the volume element, ${\cal G}(\sigma) = \exp [-\frac{1}{2}
\Delta\beta \sum_{\alpha,n}|V_\alpha| \sigma_\alpha^2(\tau_n)]$,
\begin{eqnarray}
\zeta (\sigma)={\hat{\rm T}}{\rm r}
\left[\hat{U}_\sigma(\beta,0)\right]
\label{zeta1}
\end{eqnarray}
is a trace over the one-body imaginary-time evolution operator, and
\begin{eqnarray}
\langle\hat{O}\rangle_\sigma=
\frac{\hat{\rm T}{\rm r}\left[\hat{O}\hat{U}_\sigma(\beta,0)\right]}
{\hat{\rm T}{\rm r}\left[\hat{U}_\sigma(\beta,0)\right]}.
\end{eqnarray}
In these expressions, the one-body evolution operator is defined as
\begin{eqnarray}
\hat{U}_\sigma(\tau_j,\tau_i)=
{\hat U}(\sigma(\tau_j)) \ldots {\hat U}(\sigma(\tau_{i+1}))
{\hat U}(\sigma(\tau_i))\;,
\end{eqnarray}
with
\begin{eqnarray}
{\hat U} (\sigma) = \exp\left [ -\Delta \beta \hat h(\sigma) \right],
\end{eqnarray}
and
\begin{eqnarray}
\hat h(\sigma) = \sum_{\alpha} (\epsilon_\alpha + s_\alpha
V_\alpha\sigma_\alpha) {\cal O}_\alpha.
\end{eqnarray}

In this work, we have chosen the density-decomposition described in
Ref.~\cite{ref6}, where the $\hat{\cal O}_\alpha$ are linear
combinations of the proton and neutron multipole-density operators
$\rho^{KM}(a,b) = [a^\dagger_a \times {\tilde a}_b ]^{KM}$. Written
in this way, $h(\sigma)$ can be constructed from operators that act
on protons or neutrons separately, leading to separate proton and
neutron traces; i.e., $\zeta_A(\sigma)=\zeta_{Z}(\sigma)
\zeta_{N}(\sigma)$. Of course, $N_v$ and $N_s$ then generically
represent the number of valence protons or neutrons and
single-particle states.

Equation~(\ref{mceq}) expresses the expectation value of any
observable as a multi-dimensional integral whose dimensions are at
most $N_s^2\cdot N_t$, so that it must be evaluated by Monte Carlo
techniques. Towards this end, it is necessary to define a
positive-definite weight function $W(\sigma)$. In this work, we chose
$W(\sigma) = {\cal G}(\sigma) |\zeta(\sigma)|$ so that
\begin{eqnarray}
\langle\hat{O}\rangle=
\frac{\int{\cal D}[\sigma] W(\sigma)\langle \hat{O} \rangle_\sigma
\Phi(\sigma)} {\int{\cal D}[\sigma] W(\sigma)\Phi(\sigma)},
\end{eqnarray}
with the ``sign'' $\Phi(\sigma) \equiv
\zeta(\sigma)/|\zeta(\sigma)|$. The observable
$\langle\hat{O}\rangle$ is then computed by selecting an ensemble
$\{\sigma_k \}$ chosen according to distribution $W(\sigma)$, and
computing the ensemble average; i.e.,
\begin{eqnarray}
\langle \hat O \rangle =
\frac{\sum_k \langle \hat{O} \rangle_{\sigma_k}\Phi(\sigma_k)}
{\sum_k \Phi(\sigma_k)}.
\label{sum}
\end{eqnarray}
The uncertainty in the Monte Carlo result is then related to the
ensemble standard deviation (taking into account possible
correlations between the numerator and denominator \cite{ref6}). We
have chosen the $\sigma_n$ using the standard algorithm of Metropolis
{\it et al.} \cite{ref8}. Note that a potential problem in the Monte
Carlo method is that the sign should be well determined. If
$\Phi(\sigma_k)$ oscillates wildly from sample to sample, then large
errors occur in $\langle \hat O\rangle$ because of the poorly
determined numerator and denominator in Eq.~(\ref{sum}).

Some remarks about the nature of the operators in Eq.~(\ref{mceq})
and their notation are now in order. First, since $\hat h(\sigma)$ is
only a one-body operator, the evolution operator $\hat U(\sigma)$ can
be represented as a $N_s\times N_s$ matrix $\bbox{U}(\sigma)$ in the
single-particle basis: $\bbox{U}(\sigma) = \exp(-\Delta\beta
\bbox{h}(\sigma))$, where $(\bbox{h}(\sigma))_{\alpha\beta} = \langle
\alpha |\hat h(\sigma)|\beta\rangle$. Likewise, any Slater
determinant $|\psi_S\rangle$ describing $N_v$ particles can be
represented by a matrix with $N_v$ columns and $N_s$ rows. Thouless'
theorem then implies that the operation $\hat U (\sigma)
|\psi_S\rangle$ yields only a single Slater determinant, thus
averting the need to have all Slater determinants stored in order to
evaluate Eq.~(\ref{obs}). In what follows, we represent the product
of one-body evolution operators $\hat U_\sigma (\tau_j,\tau_i)$ as
\begin{eqnarray}
\bbox{U}_\sigma (\tau_j,\tau_i) = \bbox{U}(\sigma(\tau_j))\cdots
\bbox{U}(\sigma(\tau_{i+1}))
\bbox{U}(\sigma(\tau_i))\;,
\end{eqnarray}
with the implicit understanding that
$\bbox{U}_\sigma(\beta,0)=\bbox{U}_\sigma$.

Given the matrix notation for $\hat U_\sigma(\beta,0)$, the
grand-canonical trace is given by
\begin{eqnarray}
\hat {\rm T}{\rm r}_G[\hat U_\sigma(\beta,0)] = {\rm det}(\bbox{1} +
\bbox{U}_\sigma)\;.
\end{eqnarray}
The canonical trace for fixed particle number $N_v$, $\zeta_{N_v}
(\sigma)$, can be obtained from an activity expansion by noting that
\cite{ref6}
\begin{eqnarray}
{\rm det}(\bbox{1}+\lambda \bbox{U}_\sigma) &=&
\sum_{N=0}^{N_s} \lambda^N \zeta_N (\sigma) \nonumber \\
& &=\exp\left[{{\rm tr}[\ln(\bbox{1}
+\lambda\bbox{U}_\sigma)]} \right]
=\exp\left(\sum_{n=1}^{N_s}\frac{(-1)^{n-1}}{n}\lambda^n
{\rm tr}[\bbox{U}_\sigma^n]\right),
\label{fug}
\end{eqnarray}
where the symbol ``tr'' denotes a matrix trace.

\subsection{Fourier Method for the Canonical Trace}

A major drawback of the activity expansion for computing the
canonical trace is that for $N_v\approx N_s/2$ (i.e., near half
filling), $\zeta_{N_v} (\sigma)$ involves a sum of terms that are
large in magnitude and have alternating signs. In practical terms,
the activity expansion is unstable in the mid-shell region because
these terms cancel in the sum to 10--14 orders of magnitude, and
there is a loss of numerical precision in the evaluation of
Eq.~(\ref{zeta1}). In actual calculations, we have found the activity
expansion to be stable only for $N_v \leq 4$, or, using an equivalent
hole formalism, for $N_v \geq N_s-4$.

An alternative procedure for computing the canonical trace is to use
Fourier extraction. Starting from the grand-canonical trace, and
defining $\phi_m=2\pi m/N_s$, we may write
\begin{eqnarray}
{\rm det}\left[\bbox{1}+e^{i\phi_m}e^{\beta\mu}
\bbox{U}_\sigma \right]=\sum_{N=0}^{N_s}
e^{i\phi_m N}e^{\beta\mu N}\zeta_N(\sigma)\;,
\end{eqnarray}
where $\mu$ is a parameter introduced to insure numerical stability
throughout the range of particles in the model space, and is given
below. Using the identity
\begin{eqnarray}
\frac{1}{N_s}\sum_{m=1}^{N_s} e^{ i\phi_m K}=\delta_{K0},
\end{eqnarray}
valid for integer $K$, we find
\begin{eqnarray}
\zeta_{N_v}(\sigma)=\frac{1}{N_s}\sum_{m=1}^{N_s}
e^{-i\phi_m N_v}e^{-\beta\mu N_v}
{\rm
det}\left[\bbox{1}+e^{i\phi_m}e^{\beta\mu}\bbox{U}_\sigma\right].
\label{zeta}
\end{eqnarray}
The expectation values of the one- and two-body density operators can
be computed in a similar fashion:
\begin{eqnarray}
\langle a_\alpha^\dagger a_\beta \rangle_{\sigma,N_v}=
\frac{1}{N_s\zeta_{N_v}(\sigma)}
\sum_{m=1}^{N_s}e^{-iN_v\phi_m}e^{-\beta N_v\mu}
\eta_m(\sigma)\gamma^m_{\alpha\beta}(\sigma),
\label{one}
\end{eqnarray}
and
\begin{eqnarray}
\langle a_\alpha^\dagger a_\beta a_\gamma^\dagger a_\delta
\rangle_{\sigma,N_v} &=&
\frac{1}{N_s\zeta_{N_v}(\sigma)}\sum_{m=1}^{N_s}e^{-iN_v\phi_m}
e^{-\beta N_v \mu}\eta_m(\sigma)
\nonumber \\
&
&\times\left[\gamma_{\alpha\beta}^m(\sigma)\gamma_{\gamma\delta}^m(\s
igma)
-\gamma_{\alpha\delta}^m(\sigma)\gamma_{\gamma\beta}^m(\sigma)+
\delta_{\beta\gamma}\gamma_{\alpha\delta}^m(\sigma)\right],
\label{two}
\end{eqnarray}
where
\begin{eqnarray}
\eta_m(\sigma)={\rm det}\left[\bbox{1}+e^{i\phi_m}e^{\beta \mu}
\bbox{U}_\sigma\right]
\end{eqnarray}
and
\begin{eqnarray}
\gamma_{\alpha\beta}^m(\sigma)=\left[(\bbox{1}+e^{i\phi_m}e^{\beta
\mu}
\bbox{U}_\sigma)^{-1}
e^{i\phi_m}e^{\beta \mu}\bbox{U}_\sigma\right]_{\beta\alpha}\;.
\end{eqnarray}

The observables in Eq.~(19--21) are, of course, independent of the
value of $\mu$ chosen.  However, as the $\zeta_n$ vary rapidly with
$N$, a good choice for $\mu$ is one for which the sum in Eq.~(17)
peaks at $N=N_v$. In order to find a good choice for $\mu$, we first
find the $N_s$ eigenvalues, $\lambda_i$ of $U_\sigma$, where
$i=1,\cdots N_s$, and $\mid\lambda_1\mid < \mid\lambda_2\mid <
\cdots$ (note that each eigenvalue has the form
$\mid\lambda_i\mid=\exp[-\beta\varepsilon_i]$). Thus, for the valence
particles, we define $\mu$ by
\begin{eqnarray}
\mid\lambda_{N_v}\lambda_{N_v+1}\mid^{1/2}=
\exp\left[-\beta\frac{{\rm Re}\varepsilon_{N_v}+{\rm Re}
\varepsilon_{N_v+1}}{2}\right]=
\exp[-\beta\mu]\;.
\end{eqnarray}
This prescription allows us to use Fourier extraction for all
even-even nuclei in both the $sd$ and $fp$ shells.

At first glance, the Fourier method appears to add substantial
computational effort since the computation of a determinant scales as
$N_s^3$ and it must be computed $N_s$ times in Eq.~(\ref{zeta}). In
fact, the computation of $\zeta_{N_v}$ can be simplified considerably
by computing the $N_s$ eigenvalues, $\lambda_i$, of
$\bbox{U}_\sigma$, in terms of which, the factor $\eta_m(\sigma)$ can
be written as
\begin{eqnarray}
\eta_m(\sigma) = \prod_{i=1}^{N_s}
(1+e^{i\phi_m}e^{\beta\mu}\lambda_i).
\end{eqnarray}
In addition, the matrix $\gamma_{\alpha\beta}^m(\sigma)$ is given by
\begin{eqnarray}
\gamma_{\alpha\beta}^m(\sigma) = \sum_\delta
P_{\beta\delta}(1+e^{i\phi_m}e^{\beta\mu}\lambda_\delta)^{-1}e^{i\phi
_m}
P_{\delta\alpha}^{-1}\;,
\end{eqnarray}
where $\bbox{P}$ is the transformation matrix associated with the
diagonalization of $\bbox{U}_\sigma$.

\section{Results}

In Refs.~\cite{ref5,ref6}, results obtained with the Monte Carlo
approach to the nuclear shell model were presented for even-even
$N=Z$ nuclei using a pairing plus multipole-multipole interaction. We
found empirically, and later proved for the zero-temperature and
grand-canonical formalisms \cite{ref6}, that under certain
conditions, the sign $\Phi(\sigma)$ in the Monte Carlo sampling is
always unity. These conditions are that in the density decomposition,
the sign of $V_\alpha$ is given by $(-1)^K$, where $K$ is the angular
momentum rank of the operator $\hat {\cal O}_\alpha$, and either
({\it i}) $N=Z$, or ({\it ii}) the number of protons and neutrons is
even. When a Lagrange multiplier $\omega$ is used to constrain the
$z$-component of the angular momentum, ${\hat J}_z$, (i.e., cranking
with $\hat{H}'=\hat{H}-\omega \hat{J}_z$), the time-reversal symmetry
used to prove these statements is broken, and the sign need not be
unity.

In this section, we further demonstrate the viability of the Monte
Carlo shell model approach by presenting calculations for $N \neq Z$
even-even and $N=Z$ odd-odd nuclei using the $1s$-$0d$ shell-model
configuration space. The interaction was chosen to be of the pairing
plus multipole form similar to that used in Ref.~\cite{ref5,ref6}
(the exact parameters can be obtained from the authors). In each of
the calculations presented, we used $\Delta\beta=0.0625$ and
approximately 2000 Monte Carlo samples were collected. The
independence of the individual samples was tested by computing the
auto-correlation function for $\langle H \rangle_\sigma$. All the
calculations presented here were performed on the Intel Touchstone
Gamma and Delta systems operated by Caltech for the Concurrent
Supercomputer Consortium.

We begin with a compendium of results for the Ne isotopes. We show in
Fig.~1 the expectation values (open symbols) (a)
$\langle\hat{H}\rangle$, (b) $\langle Q^2\rangle$ (quadrupole
moment), (c) $\langle J^2\rangle$ (angular momentum), and (d)
$\langle T^2\rangle$ (isospin) as functions of $\beta$ for even-even
Ne isotopes. Exact calculations within the canonical ensemble for
${}^{20,22}$Ne using the eigenvalues obtained from the shell-model
code OXBASH \cite{oxbash} are indicated by the curves in the figure.
In addition, the ground-state observables are plotted using solid
symbols at $\beta=2.5$. Generally, we see that the Monte Carlo
procedure is in good agreement with the exact calculations. Shown in
Fig.~2 are the results for the same quantities for the odd-odd $N=Z$
nucleus ${}^{22}$Na.

In Ref.~\cite{ref6}, we described how the strength function, $f(E)$,
for the operator $\hat O$ can be computed using the MaxEnt technique
to perform the inverse Laplace transform on the imaginary-time
correlation function $\langle \hat O(\tau) \hat O(0)\rangle = {\hat
{\rm T}}{\rm r} [ e^{-\beta H} e^{\tau H} \hat O e^{-\tau H} \hat O
]$. For demonstration purposes, we display the results for
${}^{22}$Ne at $\beta=2.0$ (16 and 32 time slices) in Fig.~3 for: a)
the isoscalar quadrupole, $\langle Q^0(\tau)\cdot Q^0(0)\rangle$; b)
the isovector quadrupole, $\langle Q^1(\tau)\cdot Q^1(0)\rangle$; and
c) the isovector angular momentum, $\langle J^1(\tau)\cdot
J^1(0)\rangle$, in Fig.~4. Generally, the reconstructed strength
functions are in good agreement with the exact results, especially in
those cases where most of the strength is concentrated in a single
peak as for the isoscalar quadrupole. For the situation in which the
strength function is strongly fragmented, as in the case for the
isovector angular momentum, the various lines can be reconstructed
only by using many more time slices so that sufficient information in
the small $\tau$ region of the imaginary-time response function
exits. It is clear that in this case it is necessary to disentangle
several decaying exponentials with different slopes.

We may also study rotating nuclei using the cranking Hamiltonian,
$\hat{H}'=\hat{H}-\omega J_z$. This procedure has also been discussed
in Ref.~\cite{ref6}, and we indicate here how cranking affects an
$N\neq Z$ nuclei such as ${}^{22}$Ne. The systematics for cranking
${}^{22}$Ne are shown in Fig.~4, where we display
$\langle\hat{H}\rangle$, $\langle\hat{J_z}\rangle$, and the sign
$\langle\Phi\rangle$ as a function of $\beta$ and $\omega$. We find
that the sign decays rapidly as both the cranking frequency and
$\beta$ increase. The maximum $J_z$ available to ${}^{22}$Ne is 10,
and, therefore, the $\omega=2$ case can be considered as an extreme
limit.

\subsection{Nuclear Shapes}

It is of particular interest to determine the quadrupole shape of a
nucleus as function of temperature and angular momentum. It is
generally expected that some nuclei may exhibit a sudden phase
transition from a prolate to spherical shape as the temperature
increases \cite{ref9}. In addition, as the cranking frequency is
increased a transition to oblate ellipsoids is also expected.

One measure of the quadrupole deformation is the expectation value of
$Q^2$. As is illustrated in the proceeding section, $\langle
Q^2\rangle$ is considerably larger for nuclei that are expected to
exhibit prolate deformations such as ${}^{20,22,24}$Ne, and is much
smaller for spherical nuclei such as ${}^{26}$Ne; however, $\langle
Q^2\rangle$ suffers from two shortcomings. First, $Q^2$ contains a
one-body term proportional to $\langle r^4\rangle$, which is present
even for spherical nuclei, and tends to obscure the contribution due
to the nuclear deformation. In addition, $\langle Q^2\rangle$ does
not distinguish between prolate and oblate shapes.

In order to obtain a more detailed picture of the deformation, we
examine the components of the quadrupole operator $Q_{\mu} = r^2
Y_{2\mu}^*$. Note, however, that due to rotational invariance of the
Hamiltonian, the expectation value of any component $Q_{\mu}$ is
expected to vanish. On the other hand, for each Monte Carlo sample,
there exists an intrinsic frame in which it is possible to compute
the three non-zero components $Q_0'$, $Q_{2}'$, and $Q_{-2}'$ (the
prime is used to denote the intrinsic frame). The intrinsic
quadrupole moments can then be related to the standard deformation
coordinates $\beta$ and $\gamma$ \cite{ref10} by
\begin{eqnarray}
\langle Q_{0}'\rangle_\sigma &=& \frac{3}{2\pi} \sqrt{\frac{4\pi}{5}}
\langle r^2 \rangle_\sigma \beta_\sigma \cos\gamma_\sigma \nonumber\\
\langle Q_{2}'\rangle_\sigma &=& \frac{3}{2\pi} \sqrt{\frac{4\pi}{5}}
\langle r^2 \rangle_\sigma \frac{\beta_\sigma}{\sqrt{2}}
\sin\gamma_\sigma \nonumber\\
\langle Q_{-2}'\rangle_\sigma &=& \frac{3}{2\pi}
\sqrt{\frac{4\pi}{5}}
\langle r^2 \rangle_\sigma \frac{\beta_\sigma}{\sqrt{2}}
\sin\gamma_\sigma \label{qint}.
\end{eqnarray}

The task remains to compute the quadrupole moments in the intrinsic
frame for each Monte Carlo sample. This is accomplished by computing
and diagonalizing the expectation value of the cartesian quadrupole
tensor $Q_{ij} = 3x_ix_j - \delta_{ij}r^2$ for each Monte Carlo
sample. From the three eigenvalues, it is straightforward to
determine the deformation parameters as \cite{ref11}
\begin{eqnarray}
\langle Q_{11}'\rangle_\sigma &=& \sqrt{\frac{2\pi}{5}}
\left(\sqrt{3}(\langle Q_{2}'\rangle_\sigma +\langle
Q_{-2}'\rangle_\sigma)
-\sqrt{2}\langle Q_{0}'\rangle_\sigma \right)\nonumber \\
\langle Q_{22}'\rangle_\sigma &=& \sqrt{\frac{2\pi}{5}}
\left(-\sqrt{3}(\langle Q_{2}'\rangle_\sigma +\langle
Q_{-2}'\rangle_\sigma)
-\sqrt{2}\langle Q_{0}'\rangle_\sigma\right) \nonumber \\
\langle Q_{33}'\rangle_\sigma &=& 2\sqrt{\frac{4\pi}{5}}
\langle Q_{0}'\rangle_\sigma.
\end{eqnarray}
Note, from Eq.~(\ref{qint}) one finds $Q_{22}' \leq Q_{11}' \leq
Q_{33}'$.

To illustrate the intrinsic deformation, we plot in Fig.~5 the
distribution function $f(\beta,\gamma)\beta^4\sin(3\gamma)$ for
temperatures $T=0.5$, 1.0, 2.0, and 4.0~MeV. The distribution
function was computed from the set of the $\beta$ and $\gamma$ values
from each Monte Carlo sampled, and then smoothed with a symmetric
Gaussian with a width of 0.01. Although the volume element
$\beta^4\sin(3\gamma)$ tends to push the function towards $\gamma=\pi
/6$, there is a definite trend from a prolate deformation at low
temperature to a symmetric spherical shape at higher temperatures.

\section{Conclusions}

We have further demonstrated the utility of using path integral
methods in the nuclear shell-model. We have used a realistic pairing
plus multipole interaction for $sd$-shell nuclei, and have
demonstrated systematics in the neon system. In order to evaluate
mid-shell quantities in the canonical formalism, we introduced the
Fourier extraction method. We have also indicated how shape changes
in nuclei may be calculated and observed.

In future work (in progress) we will study shape changes and behavior
of the rare-earth nuclei using the pairing plus multipole
interactions. These calculations, the largest to date, have nearly
100,000 fields over which to integrate. We also wish to further
investigate the `minus sign' problem inherent in these calculations
when certain interaction schemes are used.

\acknowledgments
This work was supported in part by the National Science Foundation
(Grants No. PHY90-13248 and PHY91-15574). D.~J.~Dean and
C.~W.~Johnson acknowledge Caltech Divisional postdoctoral
fellowships, and W.~E.~Ormand acknowledges the DuBridge postdoctoral
fellowship.

\newpage
\begin{center}
FIGURE CAPTIONS
\end{center}

\quad{FIG 1.}{The results of the Monte Carlo calculation for the
expectation values of (a) $\langle\hat{H}\rangle$, (b) $\langle
q^2\rangle$, (c) $\langle J^2\rangle$, and (d) $\langle T^2\rangle$
as a function of $\beta$ for ${}^{20}$Ne (circles), ${}^{22}$Ne
(squares), ${}^{24}$Ne (diamonds), and ${}^{26}$Ne (triangles). Where
absent, the error bars are smaller than the size of the symbols. The
solid (${}^{20}$Ne) and dashed-dot (${}^{22}$Ne) lines indicate the
canonical results obtained from the eigenvalues of an exact
diagonalization. The ground-state expectation value for each nucleus
is plotted with a solid symbol at $\beta=2.5$ (except for $J^2$, for
which the ground-state value is zero for all nuclei).}
\medskip

\quad{FIG 2.}{Same as Fig.~1 for ${}^{22}$Na.}
\medskip

\quad{FIG 3.}{Response functions for ${}^{22}$Ne are shown, along
with MaxEnt reconstruction of the strength functions. The calculated
response functions (left) of the isoscalar quadrupole (top),
isovector quadrupole (middle), and isovector angular momentum
(bottom) are shown for $\Delta\beta=0.125$ (circles), and
$\Delta\beta=0.0625$ (triangles). MaxEnt reconstruction of the
strength functions are also given for $\Delta\beta=0.125$ (dotted
line) and $\Delta\beta=0.0625$ (dashed line). Exact results given as
the delta function peaks.}
\medskip

\quad{FIG 4.}{Cranked results for ${}^{22}$Ne are given as a function
of the cranking frequency $\omega$. Calculations were performed at
$\beta=0.5$ (circles), 1.0 (squares), 1.5 (diamonds), and 2.0
(triangles).}
\medskip

\quad{FIG 5.}{Distribution functions $F(\beta,\gamma)$ are shown for
${}^{22}$Ne at different temperatures $T$.}

\end{document}